\documentclass[11pt, reqno]{article}

\usepackage{jheppub}
\usepackage{amssymb}
\usepackage{amsmath}
\usepackage[usenames,dvipsnames]{xcolor}
\usepackage{epsfig}
\usepackage{dcolumn}
\usepackage{tikz}
\usetikzlibrary{shapes.geometric, arrows}
\usepackage{upgreek}
\usepackage{setspace}
\usepackage{enumitem}
\usepackage{array,multirow,bigdelim,arydshln}
\usepackage{appendix}
\usepackage{xparse}
\usepackage[utf8]{inputenc}
\hypersetup{
	colorlinks,
	urlcolor=Maroon,
	linkcolor=Maroon,
	citecolor=Maroon
}

\NewDocumentCommand{\binomial}{omm}
 {%
  \genfrac(){0pt}{}{#2}{#3}%
  \IfValueT{#1}{_{\!#1}}%
 }
\NewDocumentCommand{\eulerian}{omm}
 {%
  \genfrac<>{0pt}{}{#2}{#3}%
  \IfValueT{#1}{_{\!#1}}%
 }

\def \s {\sigma}

\usepackage{latexsym}
\usepackage{tikz}

\title{Combinatorial Factorization}
\author{Freddy Cachazo}

\affiliation{Perimeter Institute for Theoretical Physics, Waterloo, ON N2L 2Y5, Canada}
\emailAdd{fcachazo@pitp.ca}

\abstract{The simplest integrands in the CHY formulation of scattering amplitudes are constructed using the so-called Parke-Taylor functions. Parke-Taylor functions also turn out to belong to a large class of rational functions known as MHV leading singularities. In fact, Parke-Taylor functions correspond to planar MHV leading singularities. In this note we study the behavior of CHY integrands constructed using non-planar MHV leading singularities under collinear and multi-particle factorization limits. General $n$-particle MHV leading singularities are completely characterized by a set of $(n-2)$ triples of particle labels. We give a simple operation on this combinatorial data which ``factors" the list into two sets of triples defining two lower point MHV leading singularities. The fact that general MHV leading singularities form a closed set under ``multi-particle factorizations" is surprising from their gauge theoretic origin.}

\begin{document}
\maketitle
\addtocontents{toc}{\protect\setcounter{tocdepth}{1}}
\def \tr {\nonumber\\}
\def \la  {\langle}
\def \ra {\rangle}
\def\hset{\texttt{h}}
\def\gset{\texttt{g}}
\def\sset{\texttt{s}}
\def \be {\begin{equation}}
\def \ee {\end{equation}}
\def \ba {\begin{eqnarray}}
\def \ea {\end{eqnarray}}
\def \k {\kappa}
\def \h {\hbar}
\def \r {\rho}
\def \l {\lambda}
\def \be {\begin{equation}}
\def \en {\end{equation}}
\def \bes {\begin{eqnarray}}
\def \ens {\end{eqnarray}}
\def \red {\color{Maroon}}
\def \pt {{\rm PT}}
\def \s {\sigma}
\def \ls {{\rm LS}}
\def \ma {\Upsilon}

\numberwithin{equation}{section}

\section{Introduction}

Parke-Taylor functions \cite{Parke:1986gb} are rational functions of $n$ complex variables, $\{\s_1,\s_2,\ldots ,\s_n\}$, and depend on the choice of a planar ordering. In the canonical ordering
\be
\pt(1,2,\ldots ,n) = \frac{1}{(\s_{1}-\s_2)(\s_{2}-\s_{3})\cdots (\s_{n-1}-\s_{n})(\s_{n}-\s_{1})}.
\ee
Clearly a Parke-Taylor function is cyclic invariant with respect to its planar order.

These functions are part of a family of rational functions known as MHV leading singularities \cite{Arkani-Hamed:2014bca}. In the gauge theory context, there are leading singularities in every ${\rm N}^{(k-2)}{\rm MHV}$ sector \cite{Cachazo:2008vp}. In general, leading singularities are functions of spinors $\{ \lambda_a,\tilde\lambda_a\}$ defined for each particle. The $k=2$ sector is special in that the $\tilde\lambda_a$ dependence drops out. MHV leading singularities are only functions of the Lorentz invariants $\langle \lambda_a,\lambda_b\rangle$. Writing the spinors in inhomogenous coordinates, $\lambda_a = t_a(1,\s_a)$, and factoring out the scale dependence gives rise to the functions of interested for this work. We will simply refer to these rational functions as leading singularity (or LS) functions.

Every $n$-particle leading singularity function is completely determined by a collection of $n-2$ triples of particle labels \cite{Arkani-Hamed:2014bca}. However, several collections of triples can give rise to the same LS function. For example, $\pt(1,2,3,4)$ is given either by $\{(1,2,3),(3,4,1)\}$ or $\{(2,3,4),(4,1,2)\}$. The identification of these two sets of triples is known as the square identity \cite{Arkani-Hamed:2014bca,ArkaniHamed:2012nw,Postnikov:2006kva}. Sets of $n-2$ triples modulo square identities are in one-to-one correspondence with LS functions. Let us denote the corresponding rational function associated with a set of $n-2$ triples, $T$, by $\ls(T)$. In the example above, $\pt(1,2,3,4) = \ls(\{(1,2,3),(3,4,1)\})$.

In this note we study the behavior of a general function $\ls(T)$ as one approaches a boundary of the moduli space of $n$-punctured spheres. Physically, this corresponds to a collinear or multi-particle factorization limit when $\ls(T)$ is taken to be part of the integrand of a Witten-RSV formula \cite{Witten:2003nn,Roiban:2004yf} or a Cachazo-He-Yuan (CHY) integral \cite{Cachazo:2013hca,Cachazo:2013iea,Cachazo:2014xea}. Very recently, Gao, He, and Zhang \cite{Gao:2017dek} generalized Parke-Taylor functions as CHY integrands to a family of integrands defined by tree graphs which they called Cayley integrands. In \cite{Gao:2017dek} it was proven that CHY integrals of products of Cayley functions give rise to certain sums over Feynman diagrams in a cubic scalar theory all with unit coefficients. Moreover, it was found that Cayley functions are a special class of LS functions.

Our main result is a simple and precise combinatorial description of the factorization of $\ls(T)$ purely in terms of the data in the list $T$. The process starts by separating particle labels into two sets, ``Left" ($L$) and ``Right" ($R$) with $n_L$ and $n_R$ elements so that $n_L+n_R=n$. We find that the set $T$ of $n-2$ triples splits unambiguously into two sets $T_L$ and $T_R$ by using a majority rule, i.e., a triple belongs to $T_L$($T_R$) if most of its labels are in $L$($R$). In each new set, any label in a triple that belongs to the minority is replaced by a new label, e.g. $0$. This new label means that the sets of triples $T_L$ and $T_R$ must be interpreted as LS functions with $n_L+1$ and $n_R+1$ particles each. In order for this to be the case the number of triples in each set must be two less than the number of labels, i.e., it must be that $|T_L|=n_L-1$ and $|T_R|=n_R-1$. When this is the case the CHY integrand has a simple pole and factors, otherwise the pole disappears and the factorization is not present. Not only this prescription is straightforward but it shows that the set of all $\ls(T)$ for any number of particles is closed under factorization.

In section 2 we review the definition of leading singularity functions and some of their properties. In section 3 we give  a precise parametrization of the factorization limit as a path in the moduli space of $n$-punctured Riemann spheres. In section 4 we present the procedure for {\it combinatorial factorization} illustrated by some examples. Section 5 contains the conclusions and some future directions. Appendix A contains all the rational functions associated to the triples in the examples of section 4. Finally, the proof of the factorization formula is presented in appendix B.

\section{Leading Singularity Functions}

In 2014, Arkani-Hamed, Bourjaily, Postnikov, Trnka and the author \cite{Arkani-Hamed:2014bca} introduced a set of rational functions of $n$ complex variables associated with every choice of a list of $n-2$ triples of labels from the set $\{1,2,\ldots,n\}$. The rational function associated with
\be
T = \left\{
  \begin{array}{ccc}
    (m_{11} & m_{12} & m_{13}) \\
    (m_{21} & m_{22} & m_{23}) \\
    \vdots & \vdots & \vdots \\
    (m_{n-2,1} & m_{n-2,2} & m_{n-2,3}) \\
  \end{array}
\right.
\ee
is computed from a $n\times (n-2)$ matrix $\ma$ whose columns are labeled by particle number and rows by triples in the list $T$. The matrix has only three non-zero entries in each row and it is given by
\be
\label{mm}
\ma_{\alpha, a} = (\s_{m_{\alpha 1}}-\s_{m_{\alpha 2}}) \delta_{a,m_{\alpha 3}} + (\s_{m_{\alpha 3}}-\s_{m_{\alpha 1}}) \delta_{a,m_{\alpha 2}}+(\s_{m_{\alpha 2}}-\s_{m_{\alpha 3}}) \delta_{a,m_{\alpha 1}}.
\ee
Here $\alpha\in \{1,2\ldots ,n-2\}$ labels the triples while $a\in \{1,2,\ldots ,n\}$ is a particle label.

One of the most important properties of $\ma$ is that if $|\ma^{ab}|$ denotes the $(n-2)\times (n-2)$ minor of $\ma$ obtained by removing columns $a$ and $b$, then
\be\label{redun}
\left(\frac{|\ma^{ab}|}{\s_a-\s_b}\right)^2 = \left(\frac{|\ma^{cd}|}{\s_c-\s_d}\right)^2
\ee
for any choice of $a,b,c$ and $d$. This motivates the introduction of ``the invariant" of the matrix $\ma$ associated with $T$,
\be
I(T) = \frac{|\ma^{ab}|}{(\s_a-\s_b)}.
\ee

For any set of $(n-2)$ triples with non-vanishing invariant $I(T)$ there is an associated leading singulary function given by
\be\label{formula}
\ls(T) = \frac{I(T)^2}{\prod_{\alpha =1}^{n-2}|m_{\alpha 1},m_{\alpha 2},m_{\alpha 3}|}.
\ee
Here the Vandermonde determinants in the denominator are
\be
|m_{\alpha 1},m_{\alpha 2},m_{\alpha 3}|=(\s_{m_{\alpha 1}}-\s_{m_{\alpha 2}})(\s_{m_{\alpha 2}}-\s_{m_{\alpha 3}})(\s_{m_{\alpha 3}}-\s_{m_{\alpha 1}}).
\ee
This is the class of rational functions we study as boundaries of the moduli space of $n$-punctured spheres are approached.

There are three important properties of LS functions which are not obvious and deserve some comments.

The first is that any $\ls(T)$ has only simple poles. Note that the denominator of $\ls(T)$ in \eqref{formula} can give rise to higher order poles as a pair of labels can appear in more than one triple. Therefore, it must be that the numerator always cancels them. Moreover, when this happens the pole is completely canceled. This means that a pole is present if and only if a pair of labels appears in only one of the $n-2$ triples\footnote{Of course, this must be true for all triples that are connected via the square identity discussed in the introduction.}.

The second property is that $\ls(T)$ is covariant with weight $2$ in each variable under an $SL(2,\mathbb{C})$ action on all coordinates, i.e.,
\be
\s_a \to \frac{\alpha \s_a+\beta}{\gamma \s_a+\delta}, \qquad \ls(T)\to \prod_{b=1}^n(\gamma\s_b+\delta)^2 \;\ls(T).
\ee
%
%While it is clear how the denominator of $\ls(T)$ transforms, it is not obvious that the numerator is even covariant under $SL(2,\mathbb{C})$. The complication comes from the fact that each row of the $(n-2)\times n$ matrix is labeled by a triple, say $(abc)$, and has three non-zero entries, $\{ \s_b-\s_c, \s_c-\s_a,\s_a-\s_b\}$ on columns $a,b$ and $c$ respectively. This means that neither the columns nor the rows are homogeneous in any of the variables.

The third property is that any LS function admits an expansion in terms of Parke-Taylor functions with coefficients equal to one. The decomposition is given by
\be\label{ptdecom}
\ls(T) = \sum_{\omega \in S_{n-1}}c_\omega(T)\,\pt(\omega(1),\omega(2),\ldots ,\omega(n-1),n)
\ee
where the sum is over permutations of the first $n-1$ labels. The coefficients $c_\omega(T)\in \{0,1\}$. In order to determine $c_\omega(T)$ let us declare that a triple of labels $(a\,b\,c)$ is compatible with a Parke-Taylor ordering if after removing every other label from $(\omega(1),\omega(2),\ldots ,\omega(n-1),n)$ the triple $(a\,b\,c)$ appears in the same cyclic order. For example $(1\,3\,5)$ is compatible with $(1\,2\,3\,4\,5\,6)$ and with $(5\,2\,1\,3\,4\,6)$ but not with $(5\,2\,3\,4\,1\,6)$. Now, $c_\omega(T)$ is equal to one if all $n-2$ triples in $T$ are compatible with the ordering determined by $\omega$ and zero otherwise.

The proofs of all three statements can be found in \cite{Arkani-Hamed:2014bca} and follow from the origin of LS functions as MHV leading singularities and their $d\log$ forms.

\section{Boundaries of the Moduli Space}

The moduli space of $n$-punctured spheres has boundaries that can be described in some $SL(2,\mathbb{C})$ frame by a subset of punctures approaching the origin, or equivalently by the complement of the punctures approaching infinity. A standard blow-up procedure gives rise to two spheres sharing a point with punctures distributed accordingly. It is common to denote the subsets and the spheres as ``Left" (L) and ``Right" (R).

Let $n_L$ points be on the left and $n_R = n-n_L$ be on the right. In order to study the behavior of LS functions one introduces coordinates
\be\label{coord}
\s_a =\left\{
   \begin{array}{ll}
     \frac{s}{y_a}, & ~a\in L, \\
     \frac{x_a}{s}, & ~ a\in R.
   \end{array}
 \right.
\ee
Here $s$ is a complex parameter that controls the approach to the boundary, located at $s=0$.

It is also useful to dress LS functions with the natural measure
\be
d\mu_n =\frac{1}{{\rm vol}\; SL(2,\mathbb{C})}\prod_{a=1}^n d\sigma_a.
\ee
Here $SL(2,\mathbb{C})$ acts as Mobius transformations on the coordinates $\s$ and its volume is there in order to mod out by puncture configurations which give rise to the same complex structure and hence the same complex manifold. Moreover, from the discussion at the end of the previous section it is clear that $d\mu_n \,\ls(T_n)$ is $SL(2,\mathbb{C})$ invariant. The subscript in $T_n$ was added in order to emphasize the number of external particles.

As a boundary is reached, one gets two spheres and therefore there must be one $SL(2,\mathbb{C})$ action on each. Of course, after using the coordinates \eqref{coord} the new puncture on each sphere is fixed and therefore only two redundancies are left from each $SL(2,\mathbb{C})$. Since the original measure only had three redundancies and the new one has four, the left over redundancy shows up explicitly as a measure over the deformation parameter $ds/s$.

The factorization of the measure $d\mu_n$ has been studied in many different contexts in the literature (see e.g. \cite{Vergu:2006np,Cachazo:2012pz}). The result is simply given by
\be
d\mu_n \rightarrow s^{n_L-n_R-2}\,\frac{ds}{s}\,d\mu_L\,d\mu_R \, \prod_{a\in L}\frac{1}{y^2_a}.
\ee

The main property of LS functions we explore in this paper is that they also factor according to the rules outlined in the introductions and explained in detail in the next section.
Here we only need that the list $T$ splits into two lists $T_L\cup T_R$ and when they turn out to be valid lists of triples for $n_L+1$ and $n_R+1$ particles then
\be\label{ff}
\ls(T) \rightarrow s^{-(n_L-n_R-2)}\ls(T_L)\ls(T_R)\prod_{a\in L}y^2_a.
\ee
Combining this with the measure one finds
\be
d\mu_n \ls(T) \longrightarrow \frac{ds}{s}\, d\mu_L \ls(T_L)\, d\mu_R \ls(T_R).
\ee
The procedure for constructing $T_L$ and $T_R$ from $T$ is very simple and explained below, while the proof of \eqref{ff} is given in appendix B. Now, when $T_L$ and $T_R$ are not valid lists of triples, then $\ls(T) \rightarrow {\cal O}(s^{-(n_L-n_R-2)+1})$, i.e., it is sub-leading.

The previous discussion motivates the following definition. The factorization of a rational function $\ls(T_n)$ dressed with the measure $d\mu_n$ is given by its residue at $s=0$, i.e.,
\be
d\mu_n \ls(T) \longrightarrow \oint_{|s|=\epsilon} d\mu_n \ls(T).
\ee
Computing the residue implies that when $T_L$ and $T_R$ are not valid lists of triples the answer vanishes and the corresponding channel is not present.

\section{Combinatorial Factorization of LS Functions}

The fact that leading singularity functions are defined in terms of $n-2$ triples of numbers seems strange at first but it comes from the physics of on-shell diagrams \cite{Arkani-Hamed:2014bca}. The factorization property we study in this section gives a natural reason for the way the data is presented as such a list of triples.

Consider a factorization where $n=n_L+n_R$ and assume that the data defining $LS(T)$ is given in terms of $n + m$ triples, where $m$ is a fixed integer to be determined. In order to have a chance of producing two sets of triples that define LS functions we have to be able to split the original list of $|T| = n+m$ triples into two sets with $|T_L|= (n_L+1)+m$ and $|T_R|=(n_R+1)+m$ triples each. Recall that on both, the left and the right, spheres a new puncture is generated and hence the number of points is $(n_L+1)$ and $(n_R+1)$ respectively. Imposing that $|T| = |T_L|+|T_R|$ implies that $m=-2$ as expected.

Another feature that seems odd at first sight is that the information of LS functions is encoded in triples of labels. It turns out that an odd number of labels is needed in order to perform the splitting into left and right sets unambiguously by applying a majority rule. Below we see why three labels is special.

The combinatorial factorization is performed as follows. Let $L$ and $R$ denote the set of labels on the left and right spheres, i.e., $L\cup R = \{1,2,\ldots ,n\}$.

\begin{itemize}

\item For every triple in $T$ count the number of labels in $L$. If the number of labels in $L$ is more than one then the triple belongs to $T_L$, otherwise it belongs to $T_R$.

\item Introduce a new label for the internal puncture, say $0$. Having split the list of triples $T = T_L\cup T_R$, take every label in $T_L$ which belongs to $R$ and replace it by $0$. Likewise, take any label in $T_R$ which belongs to $L$ and replace it by $0$. This final step provides the lists of triples that define  $\ls (T_L)$ and $\ls (T_R)$.
\end{itemize}

The replacement of the minority labels by $0$ explains why three labels is special. Had we started with five labels, then it would be possible that after the replacement a 5-tuple could have ended up with two equal labels.

Let us illustrate the procedure with an example. Consider the $n=8$ LS function defined by the triples
\be
T_8=\{ (8,1,2),(8,3,4),(8,5,6),(7,2,3),(7,4,5),(7,6,1)\}
\ee
under four different factorizations. $\ls(T_8)$ has the maximal number of poles possible for $n=8$ functions. Its explicit form is given in the appendix, which also illustrates the procedure reviewed in section 2 for the construction of LS functions.

The first case is a two-particle factorization (also known as a collinear limit) where $L=\{1,2\}$: The only triple that has more than one label in $L$ is $(8,1,2)$, therefore $T_L = \{(8,1,2)\}$ while $T_R = \{(8,3,4),(8,5,6),(7,2,3),(7,4,5),(7,6,1)\}$. Next, we replace $8$ in $T_L$ by $0$ to get $T_L=(0,1,2)$. Likewise we have to replace both $1$ and $2$ with $0$ in $T_R$ to get $T_R = \{(8,3,4),(8,5,6),(7,0,3),(7,4,5),(7,6,0)\}$. Both lists are valid sets of triples and their corresponding rational functions are presented in appendix A.

The second one is a three-particle factorization where $L=\{1,2,3\}$: In this case we find two triples on the left set, $T_L = \{ (8,1,2), (7,2,3)\}$, and the rest on the right set, $T_R=\{(8,3,4),(8,5,6),(7,4,5),(7,6,1)\}$. Introducing the new puncture one gets
\be
T_L =\{ (0,1,2), (0,2,3)\}, \quad T_R = \{(8,0,4),(8,5,6),(7,4,5),(7,6,0)\}.
\ee

The third example is a four-particle factorization where $L=\{1,2,3,4\}$. Following the same procedure we find
\be
T_L =\{ (0,1,2),(0,2,3),(0,3,4)\}, \quad T_R = \{(8,5,6),(7,0,5),(7,6,0)\}.
\ee

Finally, the fourth example is a three-particle factorization with $L=\{1,2,4\}$. The two sets are
\be
T_L =\{ (0,1,2)\}, \quad T_R = \{(8,3,0),(8,5,6),(7,0,3),(7,0,5),(7,6,0)\}.
\ee
Here $|T_L|=1$ which is less than the two needed for a four-particle LS function. Also note that $|T_R|=5$ which is one more than required for a six-particle LS function. Given that $T_L$ and $T_R$ are not valid sets of triples the corresponding residue vanishes.

\section{Conclusions}

In this work we have shown that the set of all leading singularities functions is closed under collinear and multi-particle factorizations. The fact that the set of LS functions behaves nicely under collinear limits is well-understood from its gauge theory origin in the MHV sector \cite{Arkani-Hamed:2014bca}. Interestingly, such MHV origin is what makes the special behavior under multi-particle factorizations unexpected. Of course, Parke-Taylor functions are a known example of a set that is closed under factorizations and has a MHV origin. However, this phenomenon has always been seen as a consequence of the world-sheet origin of $N^{k-2}MHV$ amplitudes in formulations such as the Witten-RSV construction \cite{Witten:2003nn,Roiban:2004yf}.

Here we comment on the original motivation of this work which is still at the level of a curiosity but it might be a hint of a larger structure. Start with the standard color-ordered basis of $U(N)$ gauge theory amplitudes. In this case the part of the CHY integrand that carries the color structure is
\be
{\cal C}_n= \frac{{\rm Tr}\left(T^{a_1}T^{a_2}\cdots T^{a_n}\right)}{(\s_{1}-\s_2)(\s_2-\s_3)\cdots (\s_n-\s_1)}+\cdots
\ee
where the ellipses stand for permutations of labels $\{2,3,\ldots ,n\}$.

In the study of gauge theory or NLSM amplitudes one often assumes that $N$ is large compared to the number of particles so that the traces of product of generators do not satisfy identities. Here we want to explore exactly the opposite case. Consider the case when $N=2$. Taking the generators $(T^a)_{bc} = \epsilon_{abc}$ immediately shows that when $n$ is even the trace becomes a sum of products of Kronecker delta functions defining a {\it perfect matching} of the particles. In this case it is more natural to defined a partial amplitude not as the coefficient of a trace but as the coefficient of a perfect matching. Let us consider the $n=6$ case and rewrite ${\cal C}_6$ in terms of the new set of partial amplitudes
\be
{\cal C}_n= \delta^{a_1a_2}\delta^{a_3a_4}\delta^{a_5a_6}F_{12:34:56}(\sigma) + \cdots
\ee
where the ellipses stand for permutation of labels that produce all other inequivalent perfect matchings.

Quite surprisingly, the partial amplitude $F_{12:34:56}$ is nothing but the leading singularity function defined by the triples
\be
T_6 = \{ (1\,3\,6),(2\,3\,5),(1\,4\,5),(2\,4\,6)\}.
\ee
Using the shorthand notation $\s_{ab}=\s_a-\s_b$ it is easy to write the explicit form
\be\label{sixLS}
F_{12:34:56} = \ls(T_6) = \frac{(\s_{41}\s_{35}\s_{62}+\s_{46}\s_{32}\s_{51})^2}{\s_{13}\s_{16}\s_{36}\s_{23}\s_{35}\s_{25}
\s_{14}\s_{45}\s_{15}\s_{24}\s_{46}\s_{26}}.
\ee
This observation implies that $F_{12:34:56}$ must have good properties not only on collinear limits as its MHV origin implies but also under three-particle factorizations as this half-integrand can be used in the Witten-RSV formula or CHY formula to compute NMHV amplitudes which are known to have non-trivial three-particle factorizations.

As mentioned in the introduction, Gao, He, and Zhang \cite{Gao:2017dek} generalized the Parke-Taylor functions as CHY integrands to a family of integrands called Cayley integrands. In \cite{Gao:2017dek} the identification of polytopes associated with Cayley functions was given. As suggested by Gao, He, and Zhang, it would be interesting to find a polytope interpretation for general MHV leading singularities. The fact that they have good factorization properties strongly suggests that such a polytope exists.

Finally, recent work by Early shows connections between certain leading singularity functions and combinatorial structures known as generalized permutahedra \cite{Early:2017lku}. A very intriguing feature is that other functions which are not LS functions but are closely related can naturally appear. It would be interesting to understand the larger set of functions, their factorization properties and possible physical interpretations.

\section*{Acknowledgements}

We would like to thank Nima Arkani-Hamed, Nick Early, Song He, Sebastian Mizera, Karen Yeats, and Ellis Yuan for useful discussions. This research was supported in part by Perimeter Institute for Theoretical Physics. Research at Perimeter Institute is supported by the Government of Canada through the Department of Innovation, Science and Economic Development Canada and by the Province of Ontario through the Ministry of Research, Innovation and Science.

\renewcommand{\thefigure}{\thesection.\arabic{figure}}
\renewcommand{\thetable}{\thesection.\arabic{table}}
\appendix

\section{Examples of LS Functions}

In this appendix we provide some examples of leading singularity functions. In particular, we compute the LS functions that appear in the examples provided in section 4.

Consider the $n=8$ LS function defined by the triples
\be
T_8=\{ (8,1,2),(8,3,4),(8,5,6),(7,2,3),(7,4,5),(7,6,1)\}.
\ee
The explicit form is given by
\be
\ls(T_8) = \frac{(\s_{17}\s_{23}\s_{48}\s_{57}\s_{68}+\s_{17}\s_{45}\s_{68}\s_{72}\s_{83}+
\s_{62}\s_{72}\s_{74}\s_{83}\s_{85})^2}{\s_{81}\s_{12}\s_{28}\s_{83}\s_{34}\s_{48}\s_{85}
\s_{56}\s_{68}\s_{72}\s_{23}\s_{37}\s_{74}\s_{45}\s_{57}\s_{76}\s_{61}\s_{17}}
\ee
where the shorthand notation $\s_{ab}=\s_a-\s_b$ is used in order to keep formulas compact.

Let us now give the explicit form of the LS functions appearing in the two non-trivial factorizations studied in section 3.

The first is $L=\{1,2\}$. This gives $T_L=\{ (012) \}$ while is simply given by
\be
\ls(T_L) = \frac{1}{\s_{01}\s_{12}\s_{20}}
\ee
and $T_R=\{ (834),(856),(703),(745),(760) \}$ which gives
\be
\ls(T_R) = \frac{(\s_{34}\s_{57}\s_{68}+\s_{56}\s_{74}\s_{83})^2}{\s_{03}\s_{34}\s_{37}\s_{45}\s_{48}\s_{56}\s_{57}
\s_{60}\s_{68}\s_{74}}.
\ee

The second factorization is $L=\{1,2,3 \}$. The left LS is defined by $T_L =\{ (0,1,2), (0,2,3)\}$ which gives a simple Parke-Taylor factor
\be
\ls(T_L) = \frac{1}{\s_{01}\s_{12}\s_{23}\s_{30}}.
\ee
Likewise we have $T_R = \{(8,0,4),(8,5,6),(7,4,5),(7,6,0)\}$
\be
\ls(T_R) = \frac{(\s_{48} \s_{57} \s_{60}+\s_{45}\s_{76}\s_{80})^2}{\s_{04}\s_{07}
   \s_{45}\s_{48}\s_{56}\s_{57}\s_{60}\s_{68}\s_{74}\s_{76}\s_{80}
   \s_{85}}.
\ee

Finally, the third factorization is $L=\{1,2,3,4\}$ for which $T_L =\{ (0,1,2),(0,2,3),(0,3,4)\}$ and $T_R = \{(8,5,6),(7,0,5),(7,6,0)\}$.

The corresponding rational functions are
\be
\ls(T_L) = \frac{1}{\s_{01}\s_{12}\s_{23}\s_{34}\s_{40}}, \qquad \ls(T_R) =\frac{\s_{56}}{\s_{58}\s_{86}\s_{57}\s_{76}\s_{60}\s_{05}}.
\ee

\section{Proof of Factorization}

In this appendix we provide a proof of the factorization formula discussed in the main text. Let $L\cup R =\{1,2,\ldots ,n\}$ be the partition of interest of the set of particle labels with $|L|=n_L$, $|R|=n_R$. Consider a set of $n-2$ triples $T$ which defines a LS function. Let us separate the triples into four sets according to the number of ``left" labels they contain, i.e.,
\be\label{order}
T = \{ (LLL),\ldots ; (LLR),\ldots ;(LRR),\ldots ;(RRR)\}.
\ee
This means that $T_L$ is the union of the set of triples of the form $(LLL)$ and those of the form $(LLR)$. Likewise $T_R$ is the union of triples of the form $(LRR)$ or $(RRR)$.

Let us start by assuming that $|T_L|=n_L-1$, i.e., it has the correct number of triples to define a LS function for $n_L+1$ particles. This means $|T_R|=n_R-1$.

The matrix $\ma$ is $(n-2)\times n$ and we choose to order the $n$ labels for the columns by taking those in $L$ to be the first $n_L$ and those in $R$ to be the last $n_R$. It is always possible to relabel particles so that $L=\{1,2,\ldots,n_L\}$. The order of the triples that determines the rows is chosen according to \eqref{order}. In order to compute the LS function we have to remove two columns as explained in section 2. We choose them to be the first and the last columns. This ensures that one is from the left set and one is from the right set. Before studying the structure of the matrix let us compute the form of each possible combination of coordinates $\s_a-\s_b$ using the variables introduced in \eqref{coord} which we repeat here for the reader's convenience
\be
\s_a =\left\{
   \begin{array}{ll}
     \frac{s}{y_a}, & a\in L, \\
     \frac{x_a}{s}, & a\in R.
   \end{array}
 \right.
\ee
There are three cases to consider
\be
\s_{ab} =\left\{
   \begin{array}{cc}
     s (1/y_a-1/y_b), & a,b\in L, \\
     (x_a-x_b)/s, & a,b\in R,\\
     x_a/s - s/y_b, & a\in R,\; b\in L.
   \end{array}
 \right.
\ee
Having determined the $s$ dependence of each matrix element, we can write $\ma^{1,n}$ in block form
\be
\ma^{1,n} = \left(
  \begin{array}{cc}
    A & E \\
    B & F \\
    C & G \\
    D & H \\
  \end{array}
\right)
\ee
where $A,B,C,D$ all have $n_L-1$ columns while $E,F,G,H$ have $n_R-1$ columns. The rows of $A$ and $E$ are labeled by triples of the form $(LLL)$. This immediately implies that all the entries of $E$ are zero.
%Recall the definition of $\ma$ illustrated in the examples provided in the previous section.
Moreover, each row of $A$ has only three non-zero entries, all of the form $s(1/y_a-1/y_b)$. Likewise, the rows of the matrix $D$ and $H$ are labeled by triples of the form $(RRR)$. This means that all entries of $D$ are zero and each row of $H$ has only three non-zero entries, all of the form $(x_a-x_b)/s$.

Now consider $B$ and $F$ with rows labeled by triples of the form $(LLR)$. This means that $B$ has only two non-zero entries, all of the form $x_a/s - s/y_b$ while $F$ has a single non-zero entry which has the form $s(1/y_a-1/y_b)$. This means that when computing the determinant of $\ma^{1,n}$, the entries of $F$ give rise to sub-leading terms in $s$ compared to those of the $B$ and can then be dropped. A similar argument shows that the entries in $C$ are sub-leading to those in $G$ and can be dropped.

The conclusion so far is that the matrix has the following structure to leading order in $s$:
\be
\ma^{1,n} \sim \left(
  \begin{array}{cc}
    A & 0 \\
    B & 0 \\
    0 & G \\
    0 & H \\
  \end{array}
\right).
\ee
Noting that the matrices
\be
\left(
  \begin{array}{c}
    A  \\
    B \\
  \end{array}
\right), \qquad  \left(
  \begin{array}{c}
    G \\
     H \\
  \end{array}
\right)
\ee
are square matrices of dimension $(n_L-1)\times (n_L-1)$ and $(n_R-1)\times (n_R-1)$ respectively, the determinant of $\ma^{1n}$ factors as %
\be\label{kim}
\det \left( \ma^{1n}\right) \sim \det \left(
  \begin{array}{c}
    A  \\
    B \\
  \end{array}
\right)\; \det \left(
  \begin{array}{c}
    G \\
     H \\
  \end{array}
\right)
\ee
to leading order in $s$.

Let us consider each of the two new determinants in detail and show how they can be associated with two LS functions. Start with the second one and introduce a new variable, $x_0 =0$, for later convenience
\be
\det \left(
  \begin{array}{c}
    G \\ \hline
     H \\
  \end{array}
\right) = \det \left(
  \begin{array}{c}
    \cdots,(x_b-x_0)/s,\dots ,(x_0-x_a)/s,\cdots \\ \hline
     \cdots, (x_b-x_c)/s,\ldots , (x_c-x_a)/s,\cdots ,(x_a-x_b)/s \\
  \end{array}
\right).
\ee
On the right hand side we have shown the schematic form of one row in $G$ and one in $H$. Let us factor out $1/s$ from all the rows to get
\be
\det \left(
  \begin{array}{c}
    G \\ \hline
     H \\
  \end{array}
\right) = s^{-(n_R-1)}\det \left(
  \begin{array}{c}
    \cdots,(x_b-x_0),\dots ,(x_0-x_a),\cdots \\ \hline
     \cdots, (x_b-x_c),\ldots , (x_c-x_a),\cdots ,(x_a-x_b)\\
  \end{array}
\right).
\ee
The new matrix is exactly the one used for the computation of the invariant of the set of triples called $T_R$ in section 4. More precisely, every row that came from $G$, i.e., corresponding to $(LRR)$, gives a triple where both labels from $R$ stay the same and the label from $L$ is always replaced by $x_0$. The rows coming from $H$, i.e. $(RRR)$ give rise to the same triple of labels. Moreover, the new matrix can be thought of as coming from one with $n_R+1$ columns, $\ma_R$, and deleting two columns, i.e., $\ma^{0n}_R$. This means that we can write
\be
\det \left(
  \begin{array}{c}
    G \\ \hline
     H \\
  \end{array}
\right) = s^{-(n_R-1)} |\ma^{0n}_R|.
\ee
Next, we study the first new matrix in \eqref{kim}. Repeating the same procedure one finds
\be
\det \left(
  \begin{array}{c}
    A  \\ \hline
    B \\
  \end{array}
\right) = \det \left(
  \begin{array}{c}
   \cdots, s(1/y_b-1/y_c),\ldots , s(1/y_c-1/y_a),\cdots ,s(1/y_a-1/y_b) \\ \hline
    \cdots,x_a/s-s/y_c,\dots ,s/y_b-x_a/s,\cdots \\
       \end{array}
\right).
\ee
In this case we factor out a power of $s$ from each row and introduce a new variable $y_0 \to 0$ as $s\to 0$ by identifying $y_0 =s^2/x_a$ for any $a$. Note that this is a valid operation as none of the $x_a$ are zero. Otherwise there would be a collinear factorization with $x_0$ on top of the singularity we are studying. Applying these operations we find
\be
\det \left(
  \begin{array}{c}
    A  \\ \hline
    B \\
  \end{array}
\right) = s^{n_L-1}\det \left(
  \begin{array}{c}
   \cdots, (1/y_b-1/y_c),\ldots , (1/y_c-1/y_a),\cdots ,(1/y_a-1/y_b) \\ \hline
    \cdots, (1/y_0-1/y_c),\dots ,(1/y_b-1/y_0),\cdots \\
       \end{array}
\right).
\ee
The new matrix on the right hand side can be interpreted as that coming from the set of triples $T_L$, i.e., $\ma_L$ after removing columns $0$ and $1$, i.e., $\ma^{01}_L$. However, the puncture coordinates are not the standard ones but instead they are those obtained by an $SL(2,\mathbb{C})$ transformation $\sigma \to -1/\sigma$. Before bringing the coordinates to their standard form let us study the denominator of $\ls(T)$.

It is convenient to separate the denominator of $\ls(T)$ as the product over all triples of the form $(LLL),(LLR)$ and those of the form $(LRR),(RRR)$. Again, starting with the second set one finds
\be
s^{-3(n_R-1)}\prod_{\alpha \in T_R}|m_{\alpha 1}m_{\alpha 2}m_{\alpha 3}|_x
\ee
where the Vandermonde determinants are defined using the $x_a$ variables (this is the reason for the subscript in each Vandermonde).

The first set gives
\be
s^{3(n_L-1)}\prod_{\alpha \in T_L}|m_{\alpha 1}m_{\alpha 2}m_{\alpha 3}|_{-1/y}
\ee
where the Vandermonde determinants are defined using the $-1/y_a$ variables.

Now we are ready to combine all pieces in order to write
\be
\ls(T)\! =\! \frac{|\ma^{1n}|^2/(\s_1-\s_n)^2}{\prod_{\alpha=1}^{n-2}|m_{\alpha 1}m_{\alpha 2}m_{\alpha 3}|_\s} =
\left((1/y_0-1/y_1)^2\ls(T_L)_{-1/y}\right)
\left(\ls(T_R)_{x}(x_0-x_n)^2\right)\frac{s^{n_R-n_L}}{\s_{1n}^2}.
\ee
where $\s_{1n}=\s_1-\s_n$ and
\be
\ls(T_L)_{-1/y} =\frac{|\ma^{01}_L|^2/(1/y_0-1/y_1)^2}{\prod_{\alpha\in T_L}|m_{\alpha 1}m_{\alpha 2}m_{\alpha 3}|_{1/y}},
\ee
\be
\ls(T_R)_{x} =\frac{|\ma^{0n}_R|^2/(x_0-x_n)^2}{\prod_{\alpha\in T_R}|m_{\alpha 1}m_{\alpha 2}m_{\alpha 3}|_{x}}.
\ee
Finally, we can use the behavior of LS functions under $SL(2,\mathbb{C})$ transformations to write
\be
\ls(T_L)_{-1/y} = \left(\prod_{a\in L\cup \{0\}} y_a^2\right)\ls(T_L)_y.
\ee
Separating $y_0$ from the product and simplifying one has
\be
\ls(T) \to s^{-(n_L-n_R-2)}\ls(T_L)\ls(T_R)\prod_{a\in L}y^2_a.
\ee
This is the expected result which combines naturally with the measure $d\mu_n$.

\bibliographystyle{JHEP}
\bibliography{references1}

\providecommand{\href}[2]{#2}\begingroup\raggedright\begin{thebibliography}{10}

\bibitem{Parke:1986gb}
S.~J. Parke and T.~R. Taylor, \emph{{An Amplitude for $n$ Gluon Scattering}},
  \href{http://dx.doi.org/10.1103/PhysRevLett.56.2459}{\emph{Phys. Rev. Lett.}
  {\bf 56} (1986) 2459}.

\bibitem{Arkani-Hamed:2014bca}
N.~Arkani-Hamed, J.~L. Bourjaily, F.~Cachazo, A.~Postnikov and J.~Trnka,
  \emph{{On-Shell Structures of MHV Amplitudes Beyond the Planar Limit}},
  \href{http://dx.doi.org/10.1007/JHEP06(2015)179}{\emph{JHEP} {\bf 06} (2015)
  179}, [\href{http://arxiv.org/abs/1412.8475}{{\tt 1412.8475}}].

\bibitem{Cachazo:2008vp}
F.~Cachazo, \emph{{Sharpening The Leading Singularity}},
  \href{http://arxiv.org/abs/0803.1988}{{\tt 0803.1988}}.

\bibitem{ArkaniHamed:2012nw}
N.~Arkani-Hamed, J.~L. Bourjaily, F.~Cachazo, A.~B. Goncharov, A.~Postnikov and
  J.~Trnka, \emph{{Grassmannian Geometry of Scattering Amplitudes}}.
\newblock Cambridge University Press, 2016.

\bibitem{Postnikov:2006kva}
A.~Postnikov, \emph{{Total positivity, Grassmannians, and networks}},
  \href{http://arxiv.org/abs/math/0609764}{{\tt math/0609764}}.

\bibitem{Witten:2003nn}
E.~Witten, \emph{{Perturbative gauge theory as a string theory in twistor
  space}}, \href{http://dx.doi.org/10.1007/s00220-004-1187-3}{\emph{Commun.
  Math. Phys.} {\bf 252} (2004) 189--258},
  [\href{http://arxiv.org/abs/hep-th/0312171}{{\tt hep-th/0312171}}].

\bibitem{Roiban:2004yf}
R.~Roiban, M.~Spradlin and A.~Volovich, \emph{{On the tree level S matrix of
  Yang-Mills theory}},
  \href{http://dx.doi.org/10.1103/PhysRevD.70.026009}{\emph{Phys. Rev.} {\bf
  D70} (2004) 026009}, [\href{http://arxiv.org/abs/hep-th/0403190}{{\tt
  hep-th/0403190}}].

\bibitem{Cachazo:2013hca}
F.~Cachazo, S.~He and E.~Y. Yuan, \emph{{Scattering of Massless Particles in
  Arbitrary Dimensions}},
  \href{http://dx.doi.org/10.1103/PhysRevLett.113.171601}{\emph{Phys. Rev.
  Lett.} {\bf 113} (2014) 171601}, [\href{http://arxiv.org/abs/1307.2199}{{\tt
  1307.2199}}].

\bibitem{Cachazo:2013iea}
F.~Cachazo, S.~He and E.~Y. Yuan, \emph{{Scattering of Massless Particles:
  Scalars, Gluons and Gravitons}},
  \href{http://dx.doi.org/10.1007/JHEP07(2014)033}{\emph{JHEP} {\bf 07} (2014)
  033}, [\href{http://arxiv.org/abs/1309.0885}{{\tt 1309.0885}}].

\bibitem{Cachazo:2014xea}
F.~Cachazo, S.~He and E.~Y. Yuan, \emph{{Scattering Equations and Matrices:
  From Einstein To Yang-Mills, DBI and NLSM}},
  \href{http://dx.doi.org/10.1007/JHEP07(2015)149}{\emph{JHEP} {\bf 07} (2015)
  149}, [\href{http://arxiv.org/abs/1412.3479}{{\tt 1412.3479}}].

\bibitem{Gao:2017dek}
X.~Gao, S.~He and Y.~Zhang, \emph{{Labelled tree graphs, Feynman diagrams and
  disk integrals}},  \href{http://arxiv.org/abs/1708.08701}{{\tt 1708.08701}}.

\bibitem{Vergu:2006np}
C.~Vergu, \emph{{On the Factorisation of the Connected Prescription for
  Yang-Mills Amplitudes}},
  \href{http://dx.doi.org/10.1103/PhysRevD.75.025028}{\emph{Phys. Rev.} {\bf
  D75} (2007) 025028}, [\href{http://arxiv.org/abs/hep-th/0612250}{{\tt
  hep-th/0612250}}].

\bibitem{Cachazo:2012pz}
F.~Cachazo, L.~Mason and D.~Skinner, \emph{{Gravity in Twistor Space and its
  Grassmannian Formulation}},
  \href{http://dx.doi.org/10.3842/SIGMA.2014.051}{\emph{SIGMA} {\bf 10} (2014)
  051}, [\href{http://arxiv.org/abs/1207.4712}{{\tt 1207.4712}}].

\bibitem{Early:2017lku}
N.~Early, \emph{{Generalized Permutohedra, Scattering Amplitudes, and a Cubic
  Three-Fold}},  \href{http://arxiv.org/abs/1709.03686}{{\tt 1709.03686}}.

\end{thebibliography}\endgroup

\end{document}